\documentclass{iopart}

\usepackage{amsmath}
\usepackage{iopams}
\usepackage[dvips]{graphicx}
\usepackage[latin1]{inputenc}

\def\lta{~\raise.4ex\hbox{$<$}\llap{\lower.6ex\hbox{$\sim$}}~}
\def\gta{~\raise.4ex\hbox{$>$}\llap{\lower.6ex\hbox{$\sim$}}~}
\def\ie{{\it i.e.}~}

\begin{document}

\title{Catastrophic Phase Transitions and Early Warnings in a Spatial Ecological Model}

\author{A Fernández$^1$\footnote{\mailto{arielfer@fing.edu.uy}} and H Fort$^2$\footnote{\mailto{hugo@fisica.edu.uy}}}

\address{$^{1}$Instituto de Física, Facultad de Ingeniería,
Universidad de la República, Julio Herrera y Reissig 565, 11300 Montevideo, Uruguay.\\
$^{2}$Instituto de Física, Facultad de Ciencias, Universidad de
la República, Iguá 4225, 11400 Montevideo, Uruguay}

\begin{abstract}
Gradual changes in exploitation, nutrient loading, etc. produce shifts
between alternative stable states (ASS) in ecosystems which, quite often,
are not smooth but abrupt or catastrophic.
Early warnings of such catastrophic regime shifts are fundamental for
designing management protocols for ecosystems.
Here we study the spatial version of a popular ecological model, involving a logistically
growing single species subject to exploitation, which is known to exhibit ASS.
Spatial heterogeneity is introduced by a carrying capacity parameter varying
from cell to cell in a regular lattice. Transport of biomass among cells is included
in the form of diffusion.
We investigate whether different quantities from
statistical mechanics -like the variance, the two-point correlation function
and the patchiness- may serve as early warnings of catastrophic
phase transitions between the ASS.
In particular, we find that the patch-size distribution follows a power law when the
system is close to the catastrophic transition.
We also provide links between spatial and temporal indicators and
analyze how the interplay between diffusion and spatial heterogeneity
may affect the earliness of each of the observables.
We find that possible remedial procedures, which can be followed after these early
signals, are more effective as the diffusion becomes lower. Finally, we comment
on similarities and differences between these  catastrophic shifts and
paradigmatic thermodynamic phase transitions like the liquid-vapour change
of state for a fluid like water.
\end{abstract}

\maketitle

\section{Introduction}

Ecosystems are exposed to gradual change in external conditions such as climate,
inputs of nutrients, toxic chemicals, etc. Although it is generally assumed that these
gradual variations produce also gradual changes in the ecosystems, occasionally sudden
catastrophic regime shifts may occur. Recent examples of ecosystems illustrating such
changes are the shift in Caribbean coral reefs \cite{Mc99,Ny00}, shallow lakes that become overgrown
by floating plants \cite{Sc03}, savannahs that are encroached suddenly by bushes
\cite{Lu97,Wa93} and lakes that shift from clear to turbid \cite{Sc98,Ca99}.
A simple explanation for such drastic shifts is that the ecosystem has alternative stable
states (ASS) \cite{Sc01,Carp01}. In other words, under the same external conditions the system can be in two
or more stable states. Hence, when subjected to a slowly changing external factor
(such as climate), an ecosystem may show little change until it reaches a critical point where a
sudden shift to an alternative contrasting state occurs. The presence of ASS
implies that if a system has gone through such a state shift, it tends
to remain in the new state until the control variable is changed back to a much lower level.
This hysteresis phenomenon of "history dependent" alternative equilibrium states
is well known in physics.

The simplest models for describing alternative states in ecosystems correspond to what are known in physics parlance
as \emph{mean-field} (MF) models. Neglecting all spatial heterogeneities,
these models describe the change over time of some population that characterizes the state of the ecosystem.
These models are easy to analyze and in cases without significant heterogeneity
their predictions are not very different from those of spatial models.
However, in other cases the presence of a spatial dimension profoundly alter
population dynamics or opportunities for coexistence in the real world
\cite{St97}. In fact, the oversimplification of MF models casts doubt on whether the
occurrence of an alternative stable state could be an artifact.
Moreover, verifications and predictive power with respect to catastrophic
responses to changing environmental conditions are still scarce for spatially
extensive ecosystems.
Analysis of spatially explicit models are relevant for other reasons.
For example, to understand phenomena like clumping and spatial segregation
in plant communities \cite{Le97}. It was shown that vegetation patches,
which have been extensively studied for arid lands
\cite{Ag99}, can be approached as a pattern formation phenomenon
\cite{Kl99}-\cite{Ha01}.
It has been hypothesized that vegetation patchiness could be used as a signature of imminent
catastrophic shifts between alternative states \cite{Ri04}. Evidences that the patch-size distribution of
vegetation follows a power law were later found in arid
Mediterranean ecosystems \cite{Ke07}.  This implies that vegetation
patches were present over a wide range of size scales,
thus displaying scale invariance. It was also found that with increasing
grazing pressure, the field data revealed deviations
from power laws.  Hence, the authors proposed that this power law behaviour
may be a warning signal for the onset of desertification.
These spatial {\em early warnings} complement temporal ones like the variance
of time series introduced to detect lake eutrophication
\footnote{Eutrophication is an increase in
nutrients leading to an enhanced growth of aquatic vegetation or
phytoplankton and further effects including lack of oxygen and
severe reductions in water quality, fish, and other animal populations.} \cite{Ca06}
or the impact of pollutants \cite{Br06}.

In this work we will consider the spatial version of a general ecological model
in terms of a logistically growing species whose consumption,
loss or removal (either by grazing, predation or harvesting) is represented
by a saturation curve \cite{No75, May77}. The MF version of this model, in
terms of two parameters, is known to have ASS.
In order to take into account the spatial heterogeneity of the landscape
one of the two parameters, the {\em local} parameter, is taken as dependent
on the position. The other parameter, the {\em global} or {\em control}
parameter, is taken uniform in all the system.
Our goal is to use this framework to analyze the following questions:

\begin{enumerate}

\item \label{quest1}
 How spatial heterogeneity of the environment and diffusion of matter
and organisms affect the existence of alternative stable states.

\item \label{quest2}
Whether emergent characteristic spatial patterns are really
useful as early warnings and how they are connected with temporal signs
of catastrophic shifts.

\item \label{quest3}
The search for scaling laws underlaying spatial patterns and
self-organization.

\end{enumerate}

We will address these issues by measuring typical observables of statistical
mechanics, like the spatial variance, the two-point correlation function and
the patchiness.

This work is organized as follows. In \sref{s:MF} we review the
ecological MF model and analyze it from the point of view of Catastrophe Theory
\cite{Th75}. \Sref{s:spatial} is devoted to the methods used in this study
and the characterization of the steady states reached by the system
for static situations, \ie constant values of the uniform parameter.
In \sref{s:early} we study the dynamic case in which the uniform parameter
is changing with time. This combination of a varying global
control parameter, modelling a slowly changing stressor, and a local
parameter, describing the heterogeneity of the environment, was introduced
in the case of one-dimensional models in \cite{VN05}. Besides
addressing question (\ref{quest1}), this
allows to explore question (\ref{quest2}), namely possible spatial early
warnings and their connection with temporal ones.
The analysis of the distribution of patches, the issue (\ref{quest3}), is
accomplished for changing values of the control parameter.
The usefulness of these spatial early warnings in realistic situations and to
implement remedial actions is analyzed in \sref{s:usefulness}.
In \sref{s:comparison} we compare how these 'flags', indicating the onset of
sudden shifts, display in the ecosystem under consideration
and in thermodynamics.
The conclusions and final comments are presented in \sref{s:disc}.

\section{Mean-Field Description}\label{s:MF}
Our starting point is the population model introduced to describe grazing systems
\cite{No75} and later used
in general for several ecosystems \cite{May77} and in particular for the case
of the spruce budworm \cite{Lu78, Mu93}. It involves
a biomass density $X$ which evolves in time according to:
\begin{equation}\label{noy_con_p}
\frac{dX}{dt}=rX\left(1-\frac{X}{K}\right)-\frac{cX^{2}}{h^2+X^{2}}
\end{equation}
where $r$ is the intrinsic per capita growth rate, $K$ is the carrying capacity or the number of individuals which can be supported in a given
area within natural resource limits, $c$ is the maximum consumption rate and $h$ is a half-saturation constant i.e. it
corresponds to the value of $X$ such that the effective consumption is half
of the maximum consumption rate. We can
rewrite \eqref{noy_con_p} in terms of non-dimensional quantities: $t'=rt$, $X'=X/h$, $K'=K/h$ and $c'=c/(hr)$, as
\begin{equation}\label{mean-field}
\frac{dX'}{dt'}=X'\left(1-\frac{X'}{K'}\right)-c'\frac{X'^{2}}{1+X'^{2}}
\end{equation}
In what follows, for simplicity, we will omit the $'$  for the non-dimensional variables. The r.h.s. of \eqref{mean-field} may be thought as the gradient of a potential $V$ associated to the problem:
\begin{equation}\label{potential}
V=-\int dX \left[X\left(1-\frac{X}{K}\right)-c\frac{X^{2}}{1+X^{2}}\right]
=-\frac{X^2}{2}+\frac{X^3}{3K}+c\left(X-\arctan X \right)
\end{equation}
so the equilibria correspond to the roots of the first derivative of $V$:
\begin{equation}\label{equilibria}
X\left(1-\frac{X}{K}\right)-c\frac{X^{2}}{1+X^{2}}=0
\end{equation}

\begin{figure}[htp]
\begin{center}
  \includegraphics[width=0.8\textwidth]{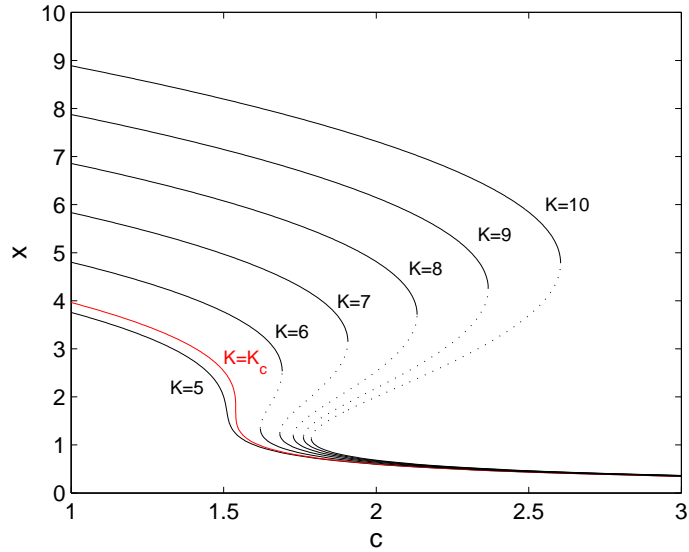}
  \caption{Folding diagrams for different values of $K$.}\label{logisticfold}
\end{center}
\end{figure}

This equation has one or three real roots (besides the trivial unstable solution $X=0$),
corresponding to one stable equilibrium state or two alternative stable states (separated by an
unstable one). It´s interesting to notice that the presence of alternative stable states is linked to the functional form assumed for the density dependent consumption. This  can be modelled by different consumption functions, which are subdivided in
three classes: linear (or Holling type I), hyperbolic (or Holling type II)
and sigmoidal (or Holling type III) \cite{Hol59}. Only for the sigmoidal
consumption there occur two stable equilibria separated by an unstable one
and therefore we have ASS.

In \fref{logisticfold} the response curve for \eqref{equilibria} is depicted for
different values of $K$. For $K\leq K_c=3^{3/2}\simeq 5.196$ only one stable solution exists for each $c$.
As long as we consider quasi stationary evolution for increasing $c$, the system would exhibit a
smooth response. On the other hand, for $K > K_c$, the response curve is folded backwards
at two \emph{saddle-node} bifurcation points. For certain values of $c$ the system can be
found either in the upper or the lower stable branch. For increasing $c$, the system starts
on the upper branch and varies its state smoothly until a threshold value is found, where a
catastrophic transition to the lower branch occurs. If at this point $c$ is decreased, we would
not be able to recover the state of the system before the transition. Instead, the system would
remain on the lower branch, until we decrease $c$ enough to reach another threshold value and 'jump'
to the upper branch. From a ecological management viewpoint, it would be desirable to anticipate
these transitions.

A general formalism for treating these catastrophic regime
shifts is the {\em Elementary Catastrophe Theory} (ECT) developed by R. Thom
\cite{Th75}. However, ECT works for static and homogeneous (MF) systems,
where there is no time or spatial dependence of the potential.
To discuss dynamics or local properties, ECT must be extended by
incorporating some external assumptions.
A change of the control parameter, reflecting changes of the external
conditions, modifies the form of the potential.
Therefore, as the shape of the potential changes, an original global minimum in which
the system sits may become a metastable local minimum because other minimum
assumes a lower value, or it even may disappear.
In this case the system must jump from the original global minimum to the new one.
ECT does not tell us when, and to which minimum, the jump occurs.
The criterion which determines this is called a {\em convention}.
Before discussing conventions we need to introduce
two important sets of points in parameter space which control structural
changes of the potential.

\begin{figure}[htp]
\begin{center}
  \includegraphics[width=\textwidth]{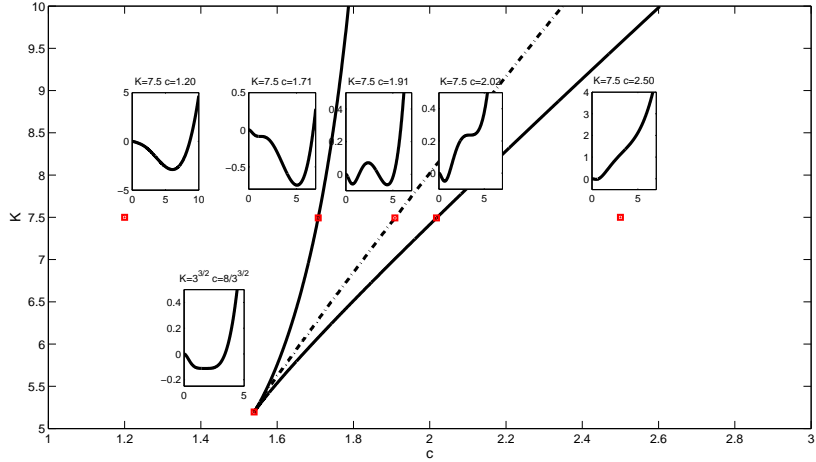}
  \caption{Bifurcation set (solid line) with a cusp point at $c=8/3^{3/2}$, $K=K_c$ and Maxwell set (dashed).
The potential $V$ is shown for selected values of $c$ and $K$.}\label{separatrix}
\end{center}
\end{figure}

The first of such sets of points is the \emph{bifurcation set} $\mathcal{S}_B$ \cite{Gil81}.
It divides the
phase space into two regions corresponding either to single stability or bistability of the system
(see figure \ref{separatrix}). For the  $(c,K)$ points on this curve the second
derivative of the potential
$V$ vanishes, so the bifurcation set is given in its parametric form by:
\begin{equation}\label{bif_parametric}
c=\frac{(x_1^2+1)^{2}}{2x_1^{3}},\quad K=\frac{2x_1^{3}}{x_1^2-1}\qquad\text{for $x_1>1$}
\end{equation}
The second set of points is called the \emph{Maxwell set} $\mathcal{S}_M$
\cite{Gil81}. On the Maxwell set the values of $V$ at two or more stable
equilibria are equal. In our case it is defined by:
\begin{gather}\label{maxwell}
\left(\frac{dV}{dX}\right)_{x_1,x_2}=0
\\ V(x_1)=V(x_2),
\end{gather}
(see the inset of $V$ for $K=7.5$, $c=1.91$ in \fref{separatrix}).

$\mathcal{S}_B$ and  $\mathcal{S}_M$ are connected to two commonly applied
criteria or conventions.
Systems which remain in the equilibrium that they are in until it
disappears are said to obey the {\em delay} convention. On the other hand,
systems which always seek a global minimum of $V$ are said to obey
the {\em Maxwell} convention.
Indeed these two conventions correspond to two extremes in a continuum of
possibilities. Furthermore, real systems may obey either of these two
conventions depending on the rate of change of the control parameters or
on other external conditions.
When the control parameters, and so the shape of $V$, change very slowly
the system tends to follow the delay convention. On the contrary, when
the control parameters change more quickly or when perturbations on the
system are big enough, the Maxwell convention describes better the dynamics
(more on this below).

\section{Spatial Model}\label{s:spatial}
A two dimensional spatial version of the previous mean-field model is given by:
\begin{equation}\label{spatial}
\frac{dX(x,y;t)}{dt}=X\left(1-\frac{X(x,y;t)}{K(x,y)}\right)-c\frac{X(x,y;t)^{2}}
{1+X(x,y;t)^{2}}+D\nabla^{2}X(x,y;t)
\end{equation}
where the carrying capacity $K(x,y)$ is a spatial heterogeneous parameter
that varies from point to point (while the parameter $c$ is taken as uniform)
and $D$ is the diffusion coefficient measuring dispersion
of $X$ in space (given in units of the intrinsic growth rate $1/r$ from \sref{s:MF}).
We simulated this model in a $L\times L$ regular square lattice, so each cell,
centred at integer coordinates $(i,j)$, can be associated with a patch of the ecosystem.
Each cell is connected to its four nearest neighbours {\it i.e.} the
von Neumann neighbourhood is used. To ensure numerical stability of the
discretization scheme even for big values of the diffusion coefficient, the
\emph{Alternating Direction Implicit Method} \cite{Pre07} is used, so evolution at each time step is divided
into two stages, treating implicitly one of the spatial coordinates at each:
\begin{gather}
\begin{split}
& (1+2\alpha)X(i,j;t+\scriptstyle\frac{1}{2}\displaystyle)-\alpha X(i+1,j;t+\scriptstyle\frac{1}{2}\displaystyle)
-\alpha X(i-1,j;t+\scriptstyle\frac{1}{2}\displaystyle) =
\\& \frac{1}{2}\left[X(i,j;t)\left(1-\frac{X(i,j;t)}{K(i,j)}\right)-c\frac{X(i,j;t)^{2}}{1+X(i,j;t)^{2}}\right]
\\& +\alpha(X(i,j+1;t)+X(i,j-1;t))+(1-2\alpha)X(i,j;t)
\end{split}\label{eq:DIM1}
\\\notag
\\ \begin{split}
&(1+2\alpha)X(i,j;t+1)-\alpha X(i+1,j;t+1)-\alpha X(i-1,j;t+1) =
\\&\frac{1}{2}\left[X(i,j;t+\scriptstyle\frac{1}{2}\displaystyle)\left(1-\frac{X(i,j;t+\scriptstyle\frac{1}{2}\displaystyle)}{K(i,j)}\right)
-c\frac{X(i,j;t+\scriptstyle\frac{1}{2}\displaystyle)^{2}}{1+X(i,j;t+\scriptstyle\frac{1}{2}\displaystyle)^{2}}\right]
\\& +\alpha(X(i,j+1;t+\scriptstyle\frac{1}{2}\displaystyle)+X(i,j-1;t+\scriptstyle\frac{1}{2}\displaystyle))+
(1-2\alpha)X(i,j;t+\scriptstyle\frac{1}{2}\displaystyle)
\end{split}\label{eq:DIM2}
\end{gather}
where $\alpha=\frac{d}{8}$ and $d$ is a reduced diffusion coefficient related with $D$ and the
lattice spacing $a$ by $d=4D/a^2$. Periodic boundary conditions ($PBC$) were used and
$L$ ranged from $100$ to $800$ (in fact, for different values of $L$ in this
range, no important differences were found). The number of time steps is typically $1000$.
Depending on the ecosystem, each time step could correspond to a day, or a
month, or a year, etc.

The range of values for the model parameters that we use are chosen to contain
the region of alternative stable states determined by the MF equations:
the carrying capacity $K(i,j)$ varies randomly from cell to cell around a
fixed spatial mean $\langle K \rangle = 7.5$ in the interval $[-\delta_K,\delta_K]$
where $\delta_K=1.0-2.5$.
Typical values for the consumption rate $c$ are between $1$ and $3$ and for
for $d$ are between $0.1$ and $5$.

\subsection{Observables}
Several quantities can be measured from the time series produced by the model:
\begin{itemize}
\item The \emph{spatial mean} $\langle X \rangle$:
\begin{equation}\label{mean}
\langle X \rangle(t)=\frac{1}{L^2}\sum_{i,j}X(i,j,t)
\end{equation}
($i$ and $j$ locate each cell of the array).

\item The \emph{spatial variance} $\sigma_X^2$:
\begin{equation}\label{variance}
\sigma_X^2=\langle X^2 \rangle-\langle X \rangle^2
\end{equation}

\item The \emph{temporal variance} $\sigma_t^2$
computed from mean values of $X$ at different times, $\bar{X}(t)$,
(here we take $\bar{X} \equiv \langle X \rangle(t)$ ) which is defined as:
\begin{equation}\label{variance_t}
\sigma_t^2=\frac{1}{\tau}\sum_{t'=t-\tau}^t \bar{X}(t')^2-\left(\frac{1}{\tau}\sum_{t'=t-\tau}^t
\bar{X}(t')\right)^2
\end{equation}
for temporal bins of size $\tau$ (typical values for $\tau$ are from 50 to
150).

\item The {\it patchiness} or {\it cluster structure}. Clusters of high (low) $X$ are defined as connected regions of cells with
$X(i,j,t)>X_m$ ($X(i,j,t)<X_m$) where $X_m$ is a threshold value.
There are different criteria to define $X_m$, one of which is stated in \sref{s:early}.

\item The \emph{two-point correlation function}
for pairs of cells at $(i_1,j_1)$ and $(i_2,j_2)$, separated a given distance $R$, which
is given by:
\begin{equation}\label{correlation}
G_2(R)=\langle X(i_1,j_1)X(i_2,j_2)\rangle -\langle X(i_1,j_1)\rangle \langle
X(i_2,j_2)\rangle
\end{equation}
\end{itemize}

\subsection{Stable states in heterogeneous media}

In this subsection we briefly describe the steady states reached by the system for
static conditions, \ie for constant values of the control parameter.
The goal is
to characterize the different alternative states, produced by eqs.
\eqref{eq:DIM1} \&  \eqref{eq:DIM2}, and  their
corresponding spatial patterns and to search for scaling laws.

\begin{figure}[htp]
\begin{center}
  \includegraphics[width=\textwidth]{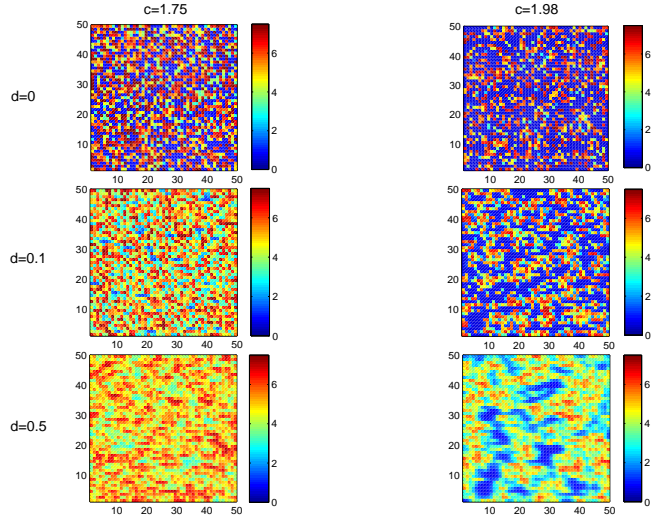}
  \caption{A portion of $50\times50$ cells from the
original $800\times800$
lattice is shown, grids representing the value taken by $X(i,j)$ at each
cell at
equilibrium for $\langle K \rangle = 7.5$, for $c=1.75$ (first column) and $c=1.98$ (second column). Each row corresponds to
$d=0$, $d=0.1$ and
 $d=0.5$ respectively.
}
\label{fig:clust&pcolor_3d_2c_k75}
\end{center}
\end{figure}

In the absence of diffusion, each cell $(i,j)$ would end up in an
equilibrium value that is completely determined by
its carrying capacity $K(i,j)$ and the initial value of $X$ at this point.
So the final state of the array
would be a random distribution of values for $X$ (see first row of \fref{fig:clust&pcolor_3d_2c_k75}).
On the other hand, dispersion among cells allows for attaining global equilibrium configurations
with some kind of spatial structure (second and third rows of \fref{fig:clust&pcolor_3d_2c_k75}).
Notice that this structure is more noticeable as $d$ increases.

\section{Alternative stable states and early warnings}\label{s:early}

Let us now study the effect of gradually increasing stress on the
system, varying $c$ from $1$ to $3$ in steps of $2/1000$. Therefore there is an
important difference with the results presented in the previous section: now
we do not let the system  "thermalize", {\it i.e.} each measure is
performed for a different value of the control parameter $c$.

We will see that some characteristics of the spatial structure may serve as
early warnings of catastrophic shifts of the system.

\subsection{Spatial and Temporal Variance}

\begin{figure}[htp]
\begin{center}
  \includegraphics[width=\textwidth]{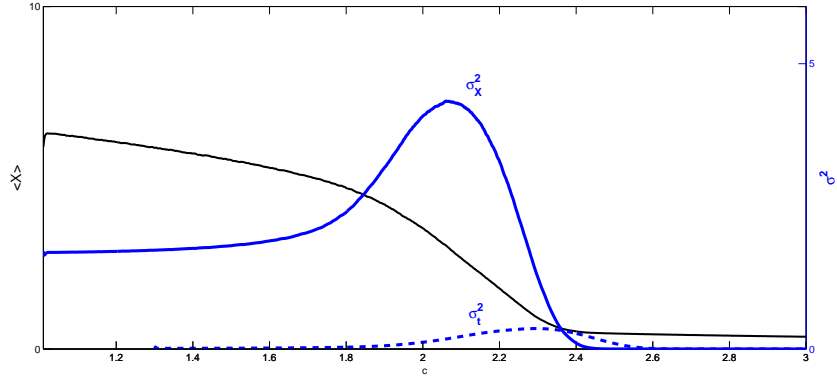}
  \caption{$\langle X \rangle$, $\sigma_X^2$ and $\sigma_t^2$
for $d=0.1$, $\langle K \rangle=7.5$.
The peak of $\sigma_X^2$ occurs at $c_m\simeq$ 2.08 and the peak of
$\sigma_t^2$ at $c\simeq 2.30$.}
\label{fig:xm&stds&stdt_k75_d01}
\end{center}
\end{figure}

In \fref{fig:xm&stds&stdt_k75_d01} we compute $\langle X \rangle$, $\sigma_X^2$ and $\sigma_t^2$ in
terms of increasing $c$ with $\langle K \rangle=7.5$, $d=0.1$ and initial condition for each $X(i,j)$ in the interval
$[0,\langle K \rangle]$. The position of peak for the spatial variance,
$c_m \simeq 2.08$, is earlier than the position of peak for the temporal
variance in nearly $110$ time steps. So $\sigma_X^2$ works better than
$\sigma_t^2$ as a warning signal for the upcoming transition.
The reason for this is clear. When estimating the temporal variance one must
consider past values in the time series, which correspond to situations
where the ecosystem is far from undergoing a transition. The spatial variance
considers only the present values, so if a signal announcing a change is
present, it is not obscured by averaging it with data where these
indications are not present.
However, notice that when the peak in $\sigma_X^2$ occurs, $\langle X \rangle$ has
already experienced a decrement of almost $50\%$ over its initial value.

\begin{figure}[htp]
\begin{center}
  \includegraphics[width=\textwidth]{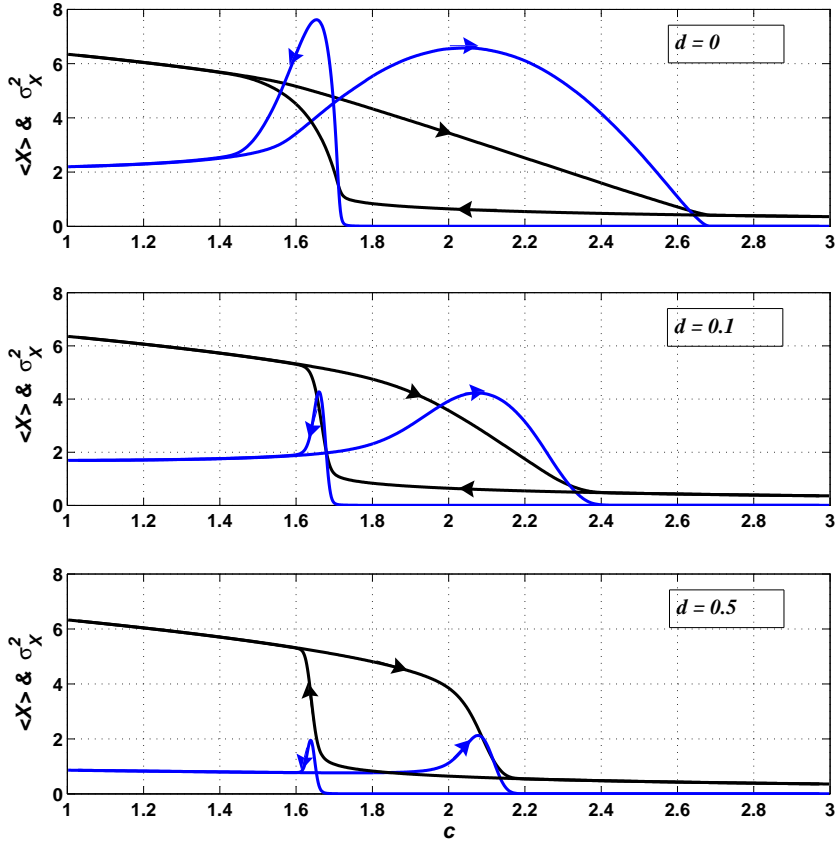}
  \caption{$\langle X \rangle$ (black curves) and $\sigma_X^2$
(blue curves) for $\langle K \rangle=7.5$ \& $\delta_K=2.5$, computed for
forward and backward changes of the control parameter $c$.
Results for $d$=0 (above), $d$=0.1 (middle) and $d$=0.5 (below).}
\label {fig:hyster-loops}
\end{center}
\end{figure}

So far we have studied the shift for increasing $c$.
Let us see what happens when $c$ is decreased.
In \fref{fig:hyster-loops} the hysteresis cycles, yielded by these
backward shifts, are shown for different values of $d$.
We observe two remarkable things. First,  the peak in
$\sigma_X^2$ is always narrower for the backward transition than in the
forward transition.
Second, the width of the hysteresis loop decreases with $d$, so diffusion
tends to make the transition more abrupt.
We will come back to discuss the effects of diffusion in greater detail
later in this section.

\subsection{Correlation}
\begin{figure}[htp]
\begin{center}
  \includegraphics[width=\textwidth]{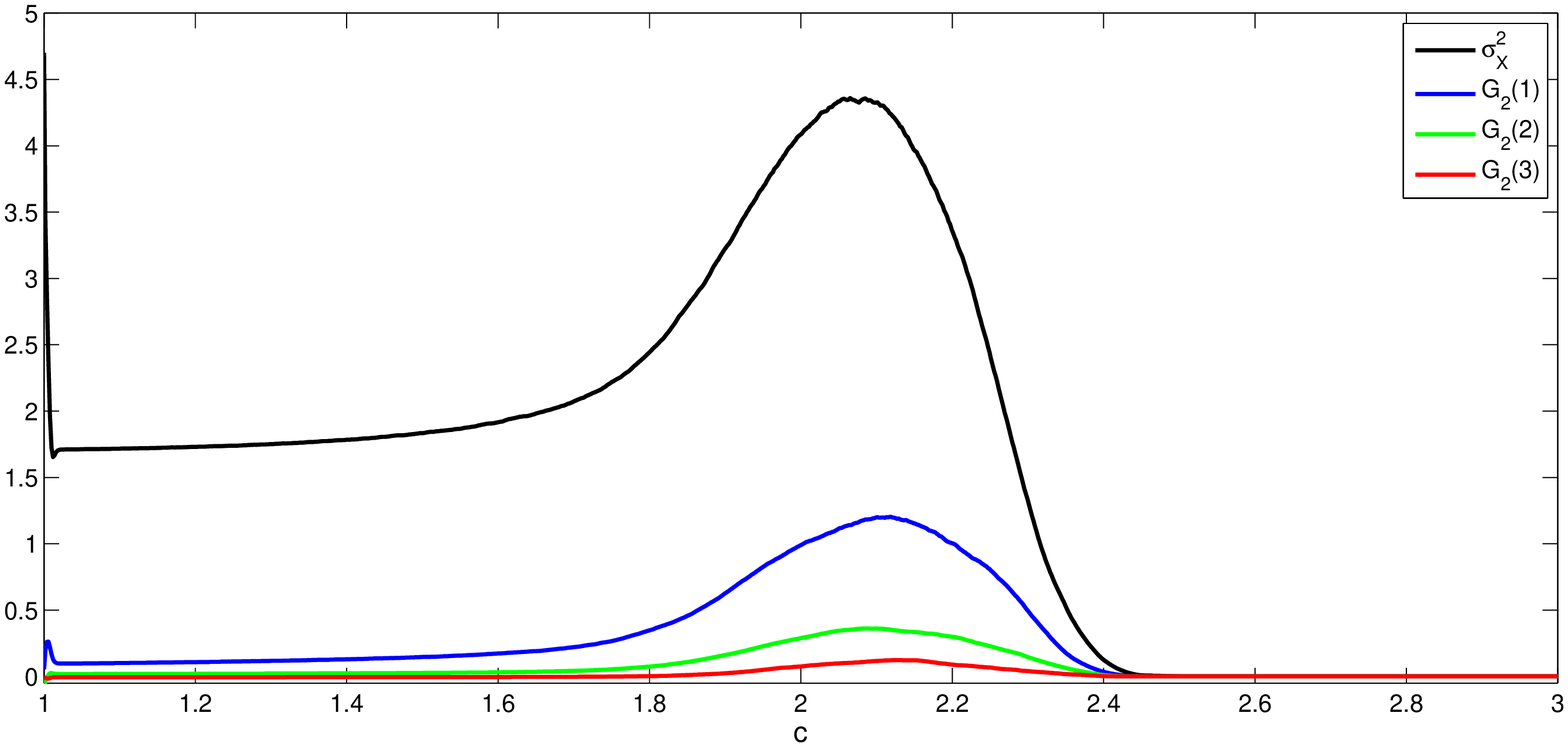}
  \caption{Two point correlation function for different lengths, $d=0.1$, $\langle k \rangle=7.5$.}\label{fig:varsG_2}
\end{center}
\end{figure}

The spatial variance is a particular case ($R=0$) of the two-point correlation function \eqref{correlation}.
We wonder if considering $R\geq$ 1 would give further information about a coming catastrophic shift.
In \fref{fig:varsG_2} the two point correlation is depicted for $R=0,1,2,3$
($R$ is measured along rows or columns of the matrix array of system's cells).
As one can see, the peak of the correlation for any $R$ occurs nearly at the same value of
the control parameter $c\approx c_m=2.08$.

\subsection{Patchiness: Cluster structure}

\begin{figure}[htp]
\begin{center}
  \includegraphics[width=\textwidth]{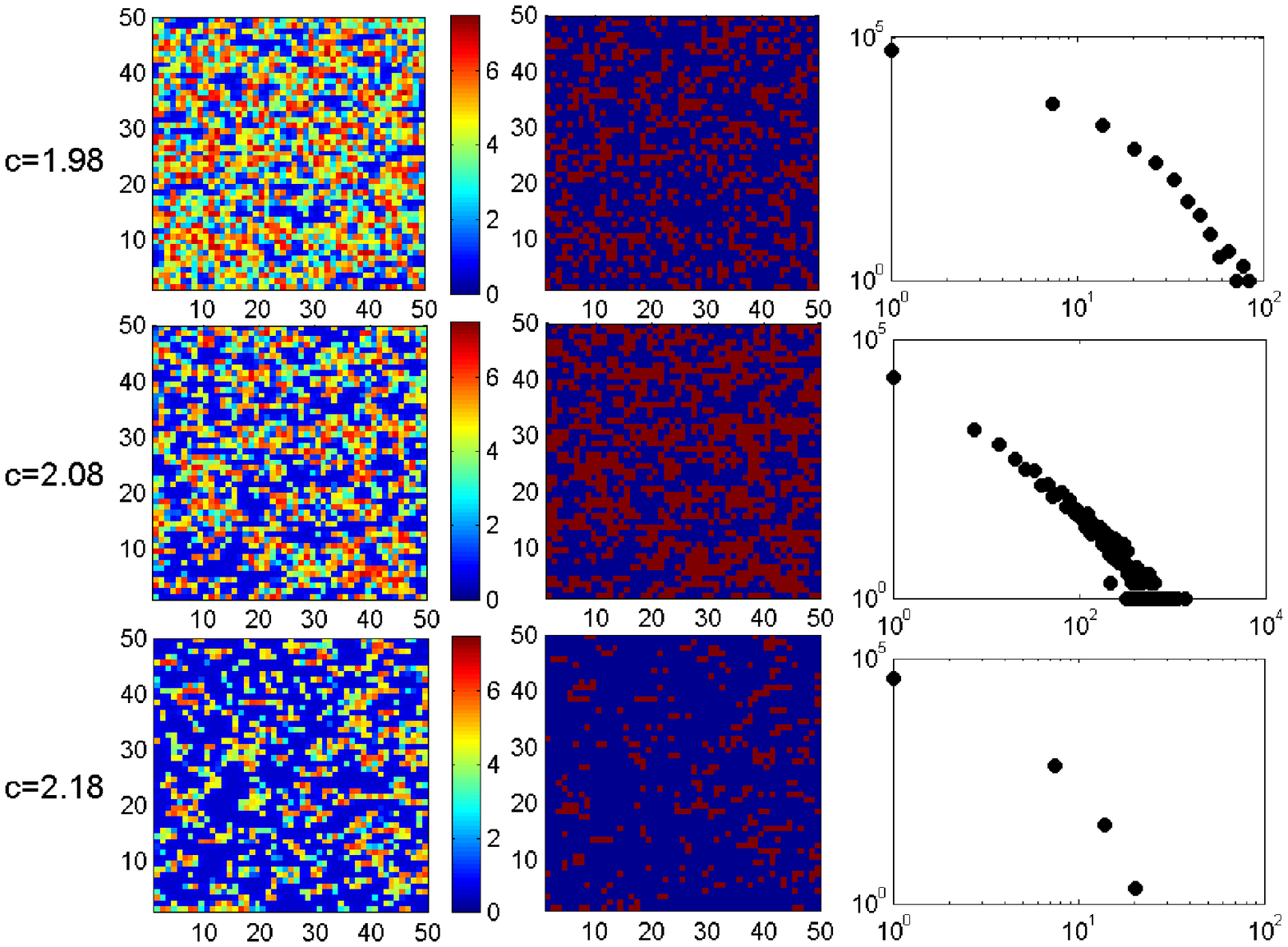}
  \caption{\emph{First column}: A portion of $50\times50$ cells from the original $800\times800$
lattice is shown, grids representing the value taken by $X(i,j)$ at each cell
for $\langle K \rangle = 7.5$, $d=0.1$. Each row corresponds to $c=1.98$, $c=2.08$
and $c=2.18$ respectively. \emph{Second column}: same as the first for binarized data.
\emph{Third column}: Number of clusters vs. area in logarithmic scale.}\label{fig:clust&pcolor_3c_k75_d01}
\end{center}
\end{figure}

In order to study the cluster structure we must define a threshold $X_m$ as a reference for the grid
values $X(i,j)$. For $\langle K \rangle = 7.5$  and $d=0.1$ the maximum in $\sigma_X^2$ is given
at $c_m \simeq 2.08$ (\fref{fig:xm&stds&stdt_k75_d01}). The value of $\langle X \rangle$ corresponding
to $c_m$ is $\langle X \rangle_{c_m}\simeq 2.89$ and we will take it as the threshold.
In the first column of \fref{fig:clust&pcolor_3c_k75_d01} we include snapshots of typical patch
configurations for $c=c_m-0.1$, $c=c_m$ and $c=c_m+0.1$ and in the second column a binary representation,
{\it i.e.} dark red (blue) cells correspond to cells for which $X>\langle X \rangle_{c_m}$ ($X<\langle X \rangle_{c_m}$).
The plots at the third column are the corresponding cluster distributions.
At $c=c_m$ the patch-size distribution follows a power law  over two decades -with exponent $\gamma\approx-1.1$ for $d=0.1$ and $\gamma\approx-0.9$ for $d=0.5$- which disappears for the smaller or greater value of $c$. Therefore this particular distribution may be considered as a signature of an upcoming catastrophic shift in the system.

\subsection{The effects of diffusion}\label{ss:diffusion}
\begin{figure}[htp]
\begin{center}
  \includegraphics[width=\textwidth]{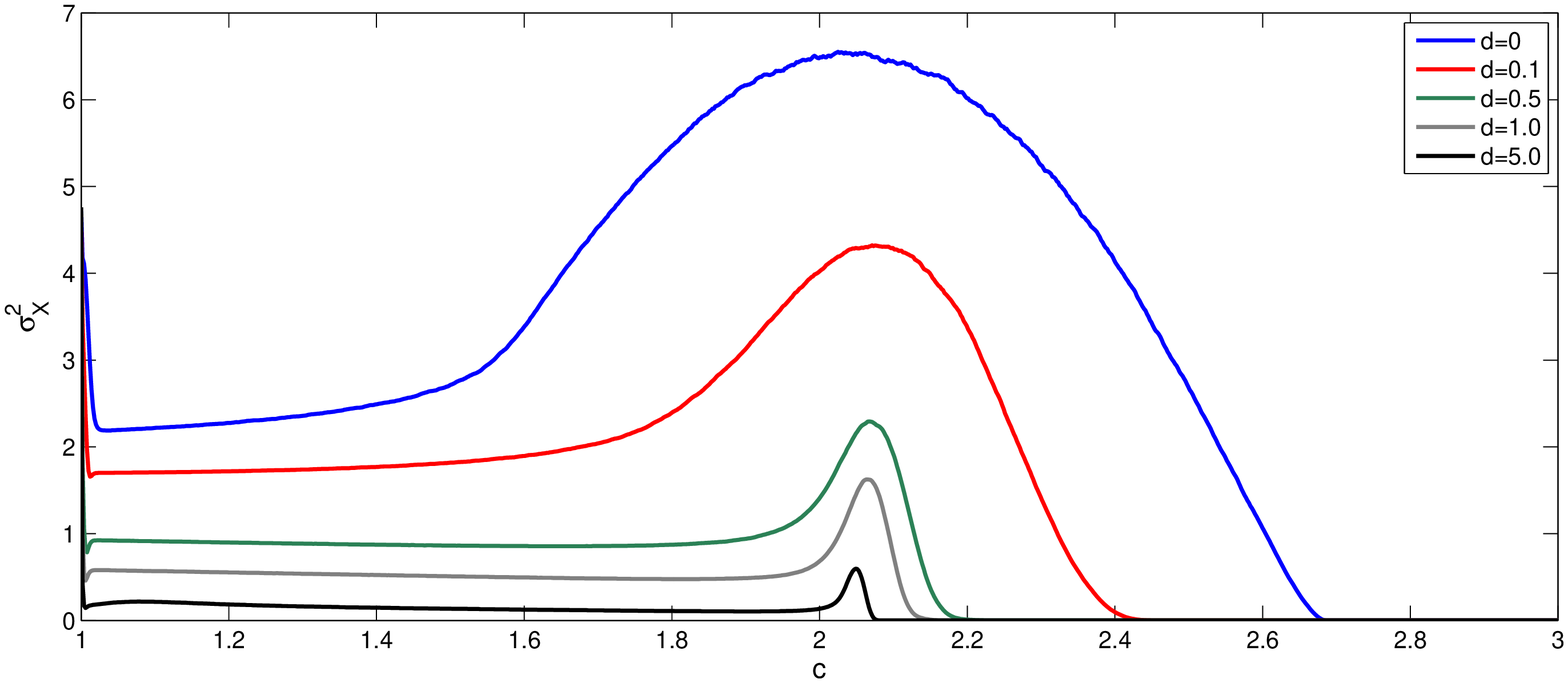}
  \caption{Spatial variance for $\langle K \rangle=7.5$ and
different values of the discrete diffusion coefficient.}
\label{fig:vars_diffd}
\end{center}
\end{figure}

\begin{figure}[htp]
\begin{center}
  \includegraphics[width=\textwidth]{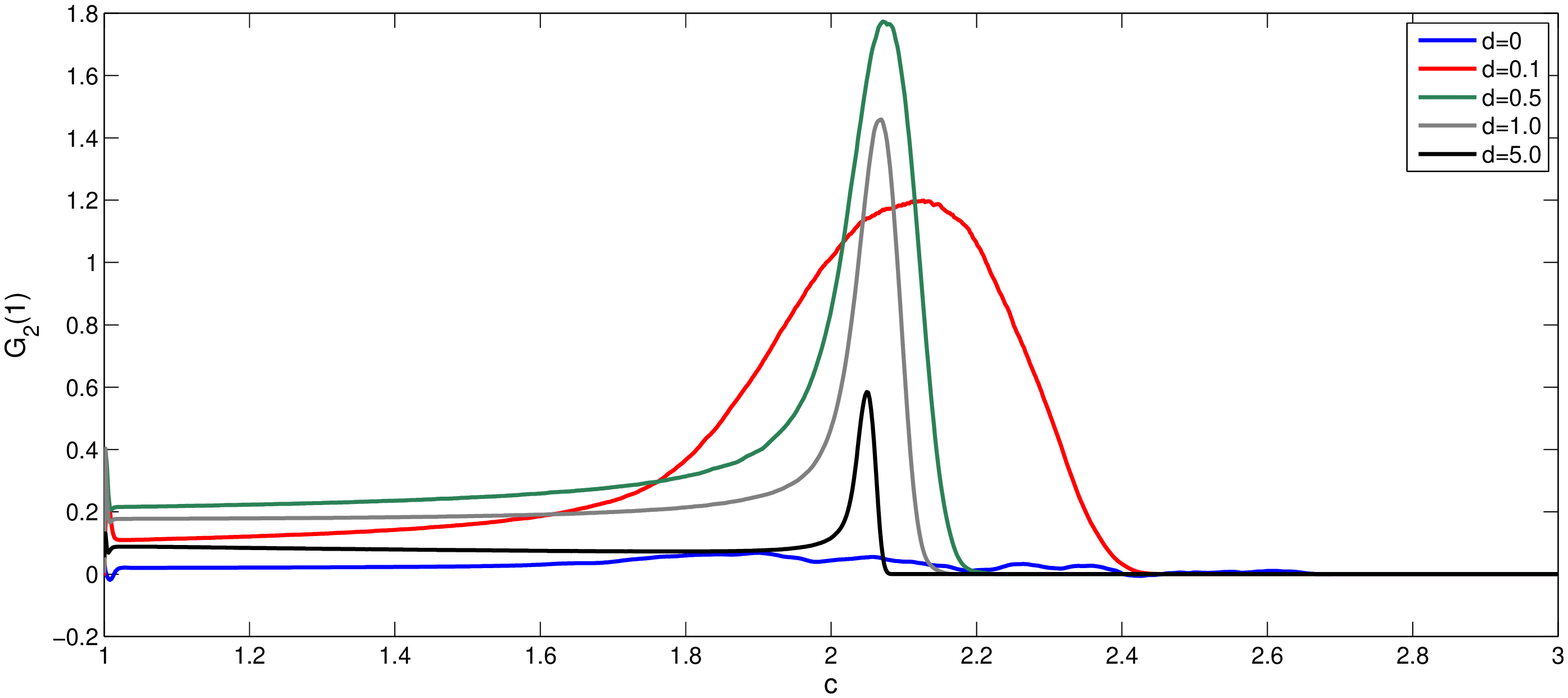}
  \caption{Two point correlation function at distance $R=1$ for different values of the discrete diffusion coefficient,
  $\langle K \rangle=7.5$.}
\label{fig:G2_diffd}
\end{center}
\end{figure}

Now we will consider the dependence on the diffusion coefficient $d$
of the different introduced spatial signals.
\Fref{fig:vars_diffd} and \fref{fig:G2_diffd} show, respectively,
the variance and correlation
for several values of $d$ between $0$ and $5.0$.
Notice that the influence of diffusion on $\sigma_X^2$ and on the
correlation is just the opposite.
In fact for $d=0$ there's almost no correlation and the peak of $\sigma_X^2$
is maximum. On the other hand, for $d\simeq0.5$ the peak in the correlation is maximum
whereas the peak for $\sigma_X^2$ is smaller and much more narrower.
This is because we have two opposite 'forces' operating over the ecosystem.
On the one hand its intrinsic underlying spatial heterogeneity
($K$ = $K(i,j)$ ) promotes spatial fluctuations between nearest neighbours,
while the diffusion term tends to smooth out these differences.

\begin{figure}[htp]
\begin{center}
  \includegraphics[width=\textwidth]{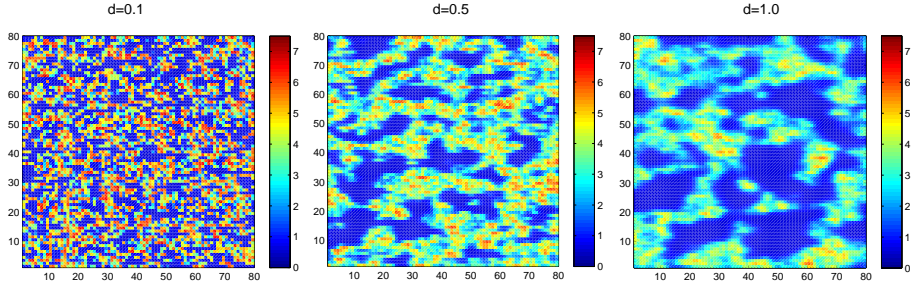}
  \caption{A coloured grid representing the value taken by $X(i,j)$ at each cell
 for the value of $c_m=2.08$.
A portion of the lattice containing 80$\times$80 cells is shown.
For $d$=0.1: $\sigma_X^2\simeq 4.315$ and $G_2(1)\simeq 1.173$.
For $d$=0.5: $\sigma_X^2\simeq 2.292$ and $G_2(1)\simeq 1.762$.
For $d$=1.0: $\sigma_X^2\simeq 1.626$ and $G_2(1)\simeq 1.458$.}
\label{fig:pcolor_diffd_maxvars}
\end{center}
\end{figure}

The resulting spatial patterns are shown in \fref{fig:pcolor_diffd_maxvars}
for $c_m$ and different values of $d$.
For low diffusion, e.g. $d=0.1$,
a typical configuration of the system consists in small patches
of arbitrary different colours (that is, large global differences,
measured by the spatial variance, and low correlation). As $d$ increases,
nearest neighbour cells group into larger patches or `supercells' of the same
`color'.
The result is a lower variance and a higher correlation. For example,
the values of $\sigma_X^2$ and $G_2(1)$ for $d = 0.1$ and $d = 0.5$
are: $\sigma_X^2\simeq 4.315$ for $d=0.1$ vs. $2.292$ for
$d=0.5$ and $G_2(1)\simeq 1.173$ for $d=0.1$ vs. $1.762$ for $d=0.5$.
Nevertheless if $d$ increases even more, the color segregation
is so strong that at this point the size of the supercells start to decrease
lowering the correlation.
So for $d=1.0$ we have: $\sigma_X^2\simeq 1.626$,
$G_2(1)\simeq 1.458$.

\section{Usefulness of the spatial early warnings}\label{s:usefulness}

To determine the usefulness of the warning indicators presented in the previous section
it is necessary 1) to assess their practicality and 2)
if they really allow the implementation of corrective actions to avoid the catastrophic
shift.

\subsection{Practical considerations: dealing with incomplete and noisy information}

\begin{figure}[htp]
\begin{center}
  \includegraphics[width=.7\textwidth]{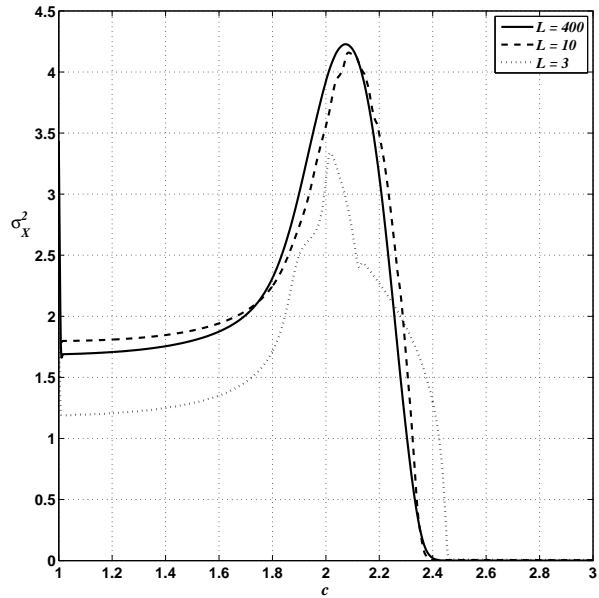}
  \caption{$\sigma_X^2$
for $d=0.1$, $\langle K \rangle=7.5$, $\delta_K=2.5$ calculated on
lattices of size $L_s$=3 (dotted line), $L_s$=10 and $L_s$=400 (the entire
lattice).}
\label{fig:var_samples}
\end{center}
\end{figure}

\begin{figure}[htp]
\begin{center}
  \includegraphics[width=\textwidth]{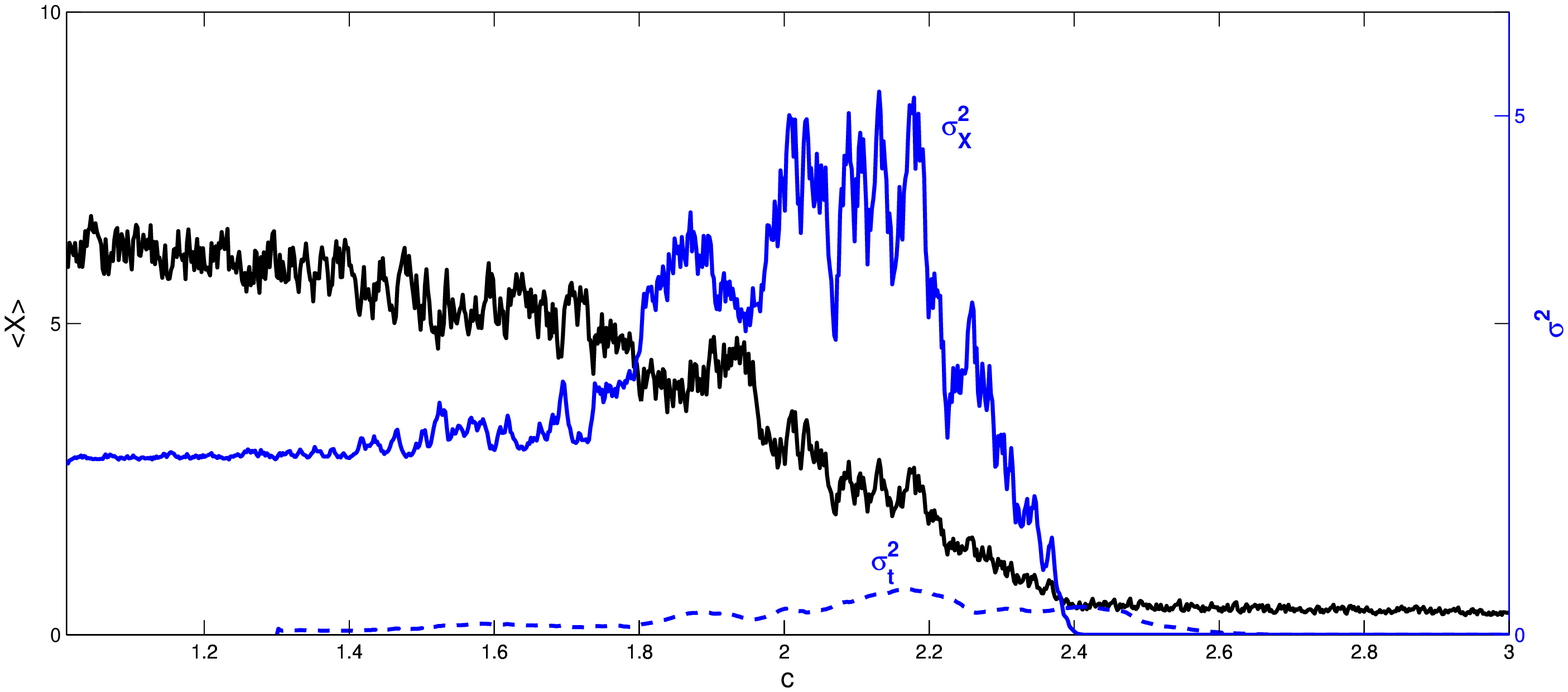}
  \caption{Same as \fref{fig:xm&stds&stdt_k75_d01} for
$c\pm 0.5$.}
\label{fig:xm&stds_noise_k75_d01}
\end{center}
\end{figure}

Calculating variances over grids consisting in a large number of sites (e.g.
400$\times$400 or 800$\times$800) is easy on a computer but involves a
formidable task from a measuring point of view. So, in order to assess the
practical difficulty of estimating $\sigma_X^2$,  we have performed
calculations over sample grids of different sizes $L_s < L$. In
\fref{fig:var_samples} we observe that the signal does not depend
qualitatively on the number of points on the grid that are considered to
estimate $\sigma_X^2$. In fact, even for a very small sample of 9 points,
$\sigma_X^2$ still exhibits a noticeable peak.
Of course, the quality of the signal improves with the size of the sample.
Additionally, since the data from real ecosystems may be very noisy, it is
worth considering how the presence of noise alters results.
So we assume some level of noise by adding to $c$ a small
random value belonging to some interval $[-\delta_c,\delta_c]$.
In \fref{fig:xm&stds_noise_k75_d01}
we show $\langle X \rangle$ and $\sigma_X^2$
for $\delta_c=0.5$.
The rise of $\sigma_X^2$ and the anticipation to the temporal variance are
still observed.

\subsection{Possible remedial actions}

\begin{figure}[htp]
\begin{center}
  \includegraphics[width=\textwidth]{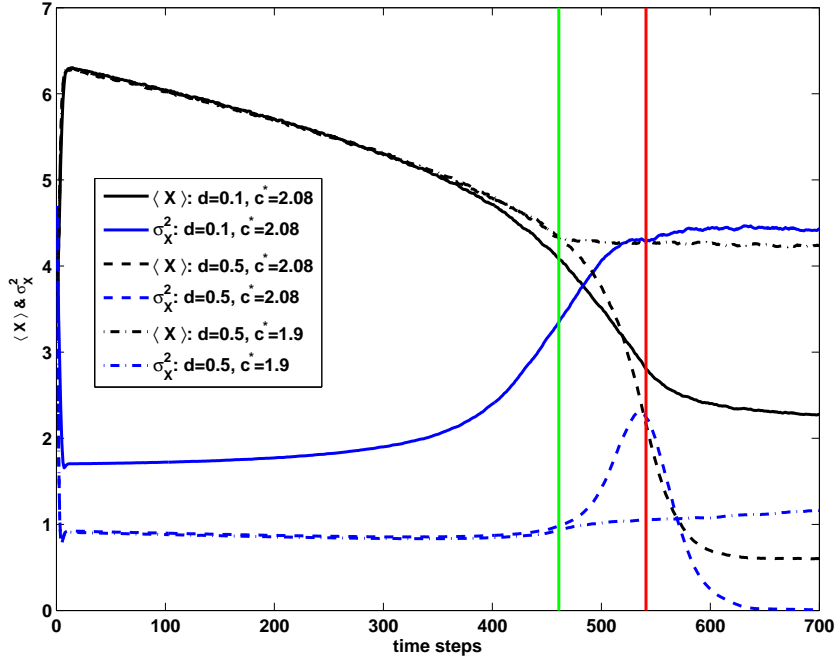}
  \caption{$\langle X \rangle$ (black) and $\sigma_X^2$ (blue) for $\langle
K \rangle=7.5$ in the case of a remedial action consisting in keeping
constant the control parameter after it reaches some threshold value $c^*$.
The red line indicates a threshold $c^*$ coinciding with the peak of
$\sigma_X^2$, $c^*$ = $c_m \simeq$  2.08. Full (dashed, dash-doted) curves correspond to
$d$=0.1 ($d$=0.5).
The green line points a value of $c^*$ before $c_m$, $c^* = 1.9$}
\label {fig:remedy}
\end{center}
\end{figure}

We will study the
consequences of a simple remedial action consisting in immediately stopping  the increase
of the control parameter after it reaches some threshold value $c^*$.
In \fref{fig:remedy} we show the effect of keeping $c$ constant to $c^*$ for
different values of $c^*$ and $d$.
For instance, if the measure is applied at the very position of the peak of
$\sigma_X^2$, $c^*$ = $c_m \simeq 2.08$ (for $\langle K \rangle=7.5$), its
usefulness depends on the value of $d$. For $d$
small ($d=0.1$) the decay in $\langle X(t)\rangle$ stabilizes soon to a
value above 2 {\it i.e.} the system remains in a mixed state. On the other
hand, for larger values of $d$ ($d=0.5$) the decay in $\langle X(t)\rangle$
continues and the ecosystem passes to the alternative state with low
biomass, $\langle X(t)\rangle$ $<$ 1.
This figure also shows that, for $d=0.5$, the remedial measure is effective
when applied before $\sigma_X^2$ reaches its maximum at $c_m$, for $c^* =
1.9$. We checked that, for moderate or high diffusion (d $\gta$ 0.5), this
recipe of management works if $c^*$ is taken between the line corresponding
to $\mathcal{S}_M$ and the right fold line of $\mathcal{S}_B$ (closer to the
first than to the second one). So a possible criterion to choose $c^*$ is as
the points belonging to $\mathcal{S}_M$.

\section{A comparison with a thermodynamic phase transition like
liquid-vapour: from the delay to the Maxwell convention}\label{s:comparison}

Catastrophes have characteristic fingerprints or 'wave flags'. Some of the
standard catastrophe flags are: {\em modality, sudden jumps, hysteresis} and
a {\em large} or {\em anomalous variance} \cite{Gil81}.  These are precisely
the signals we found for the considered spatial heterogeneous ecological
model representing a species or set of species subject to exploitation
(grazing, harvesting or predation).

It is interesting to analyze similarities and differences with the
liquid-vapour transition in a fluid, like water. Therefore, the biomass density
$X$ would correspond to the fluid density, the liquid to the high biomass
density attractor and the vapour to the low biomass density attractor. Let
us compare the above catastrophe flags for the fluid vs. the ecosystem:

\begin{itemize}

\item Modality: The fluid is bimodal in the neighbourhood of the liquid-gas
coexistence curve, having well defined liquid and gas states. So this is
similar in both systems.

\vspace{2mm}

\item Sudden Jumps: In the case of the fluid it is certainly true that
sudden jumps occur, since there is an abrupt increase in volume when a
liquid transforms into vapour. However, this large change in volume occurs
when a slight change in the temperature and pressure moves the fluid from
one side of the coexistence curve to the other. Hence, the liquid-vapour
coexistence curve can be identified with $\mathcal{S}_M$ and the water
changes of state obey in general the Maxwell convention.

On the other hand, the shift in the considered model always
obeys the delay convention: the ecosystem remains
in the higher attractor (higher values of $X$) until the bifurcation set is
completely traversed.
However, we have seen in  \sref{s:MF} that when perturbations are big enough
to allow the switching between equilibria on different
stability branches, the systems may follow the Maxwell convention.
Hence we will consider the effect of a sudden perturbation of the environment,
represented here by a sharp decrease of the average carrying capacity
$\langle K \rangle$ followed by a slow recovery.
\Fref{fig:Perturbation_and_Jump} illustrates this from a MF point of view:
$K$ is initially equal to $7.5$, and for a value of the control parameter
$c=1.68$ suddenly decreases to $6$. Afterwards $K$ increases slowly in time (as $c$ does) until it reaches
its original value just before the system crosses  $\mathcal{S}_M$ at $c=1.915$.
The insets show the shape of the potential $V(X)$
just before the perturbation and after recovery.

\begin{figure}[htp]
\begin{center}
  \includegraphics[width=\textwidth]{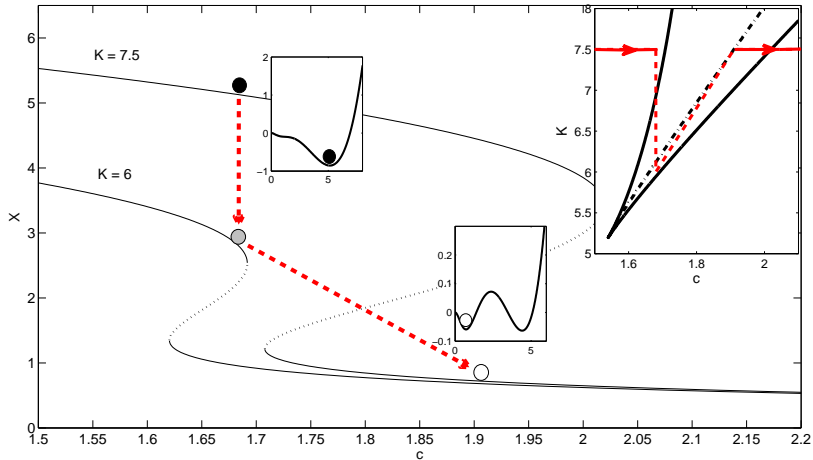}
  \caption{Variation in X produced by a global perturbation on $K$ which
suddenly decreases from $7.5$ to $6$ and slowly recovers later. The ecosystem is represented by a black
ball before the perturbation, gray ball at intermediate step and white ball after recovery. Iso-$K$ curves for
$K=7.5$ and $6$ are depicted. The red arrow represents the perturbation.
Insets show the shape of the potential $V(X)$ just before the
perturbation and after recovery. \emph{Upper right inset}: path followed by the system under perturbation
in $c-K$ phase space.
}
\label{fig:Perturbation_and_Jump}
\end{center}
\end{figure}

What happens in the case of the spatial heterogeneous and diffusive model?
In \fref{fig:Perturbation_and_Jump_xm_vs._c} we show the evolution of the
system for a completely similar perturbation in $\langle K \rangle$.
Instead of remaining close to the initial attractor (upper branch of $K=7.5$),
the system rapidly falls to the lower branch of $K$ = 6.0 (which corresponds to
the minimum value of the potential $V$). Next it approaches more slowly to the lower
branch of $K$ = 7.5 until it arrives to it for $c\simeq$ 1.915.
So one can conclude that this type of perturbation on the system produces a
change of convention: from delay to Maxwell.

\begin{figure}[htp]
\begin{center}
  \includegraphics[width=\textwidth]{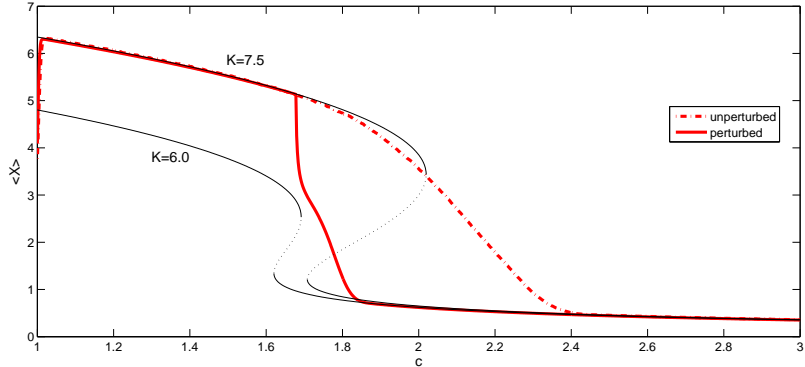}
  \caption{The effect on $\langle X \rangle$ of  a global perturbation on $\langle K \rangle$ which
suddenly decreases from  $\langle K \rangle = 7.5$ to $6$ and slowly recovers later.
Thin lines represent iso-$K$ curves for $K=7.5$ and $K=6.0$.
}
\label{fig:Perturbation_and_Jump_xm_vs._c}
\end{center}
\end{figure}

\vspace{2mm}

\item Hysteresis: In everyday situations one does not observe hysteresis in
the liquid-gas phase transition of water: the liquid usually boils at the
same temperature at which the vapour condenses. In other words, water
changes of state obey in general the Maxwell convention. Nevertheless, a
careful experimentalist can obtain an hysteresis cycle by first raising the
temperature and superheating the liquid, and after evaporation, cooling the
gas below the condensation point. Indeed the coexistence curve is surrounded
by two {\em spinodal lines} which determine  the limits to superheating and
supersaturation. These spinodal or fold lines can then be identified with
$\mathcal{S}_B$.

\vspace{2mm}

\item Anomalous Variance: When a fluid  condenses (boils) from its gas
(liquid) to its liquid (gas) state, small droplets (bubbles) are formed. As
a consequence, the variance of the volume may become large, similarly to
what happens for the ecosystem.

\end{itemize}

\section{Conclusion}\label{s:disc}

We have analyzed a spatial ecological model whose MF version has been
widely used to describe different relevant processes, ranging from pest
outbreaks to habitat desertification and harvesting of aquatic plants.
This model has alternative attractors, and is subjected
to random spatial dispersion.

For large enough values of the diffusion coefficient $d$, the system
self-organizes producing characteristic spatial patterns.

When changing the control parameter $c$, the
transition from one attractor to the other is
according to the delay convention. Nevertheless, is remarkable that,
the transition occurs according to the Maxwell convention when a large
enough perturbation is considered.
This is similar to what happens in thermodynamics. In general
we encounter the Maxwell convention in thermodynamics: water either boils
or condense when it reaches the saturation temperature.
However, it is possible, with sufficient care, to superheat water
or supercool steam, although any disturbance will produce an immediate change
of phase.

We have considered several spatial quantities both to characterize the state
of the ecosystem and to use them as early warnings.
Providing early warning signals is central both for management and recovery
strategies of ecosystems.
One of such observables is the spatial variance $\sigma_X^2$
by measuring samples of $X$ on a grid of points.
It was found that a grid containing few points might be sufficient for the
purpose of extracting an appropriate signal, and that a significant growth in
$\sigma_X^2$ could serve as an early warning of an imminent transition.
This significant growth in the spatial variance is still observed even in the
presence of moderate noise too. This is not surprising since noise, on the other hand, has been addressed as a
promoting factor over persistence of alternative stable states in other ecosystems \cite{DOd06}.

The spatial variance shows an advantage over the temporal one, as
$\sigma_X^2$ soars before than $\sigma_t^2$. The
explanation for
this is simple: since the former corresponds to a snapshot of the present
state of the system while the latter includes in its computation data for
previous times where the fluctuations were still small.

The origin of the rise in $\sigma_X$,
is tied to the emergence of spatial patterns, in the form of patches of
high/low concentration of $X$.
We then conclude that the visualization of the onset of those patches, for
example by aerial or satellite imaging, may be another indicator of the
imminence of a catastrophic shift and an effective way of
anticipating this transition. Furthermore, we found that at the very maximum
of $\sigma_X$ the distribution of sizes of patches becomes
power law, so this particular distribution could serve as an early warning.
Power law distribution has also been found in other systems as a signature of self-organization \cite{Pas02,Van08}.

Another observable of interest is the two-point correlation.
We found that as long as the diffusion coefficient $d$ increases
the peak in $\sigma_X^2$ decreases and the correlation increases.
This dependence on $d$ connected to the spatial patterns that emerge as
$d$ increases is the result of two factors that point in opposite
directions: the intrinsic spatial heterogeneity of the ecosystem
versus the dispersion or diffusion.
Therefore, for low diffusion, $\sigma_X^2$ is the most appropriate
of these two indicators to detect catastrophic shifts while the correlation
works less well. On the other hand, for high diffusion, the correlation may
become a more useful quantity to analyze.

How helpful are all these warning signals in designing effective management
protocols? Leaving aside economic considerations (which are beyond the scope
of our analysis) this depends on different factors.
For example, on the degree of diffusion (the size of $d$):
The larger the diffusion the earlier the corrective action should be taken.
For low values of $d$, a drastic measure, of immediately freezing the
consumption rate $c$, is effective even when it has reached the value $c_m$
at which the spatial variance is maximal. For larger values of $d$ the
remedial action taken at $c_m$ can no longer avoid the catastrophic shift.
Instead, provided there are no large perturbations, a simple quantitative
criterion to take the remedial action is when $c$ is over the line
$\mathcal{S}_M$. We have also seen that abrupt changes of the environmental
conditions, reflected as sudden large variations of the parameters, can
precipitate the transition to the low biomass catastrophic alternative
state.
It is worth to remark that early warning signals just provide a time where
it is still possible to act, but not at all when the situation is easily
reversible.

Of course, the quantitative details of our conclusions
depend on the choice of parameter values employed in our model.
Nevertheless, we have verified that the qualitative behaviour of our results
do not depend strongly on those values. Rather it appears that our main
conclusions should hold: spatial signals -variance, correlation and
patchiness- are earlier than the temporal variance.
Furthermore, spatial patterns formed in the process
could be the fastest detectable warning that a catastrophic change is about
to occur. Similar results have also been found for an eutrophication model
\cite{Do09} where alternative stable states are also present.

Finally, in Ecology, as far as we know, the studies focus either on spatial
early warnings (for instance refs. \cite{Ri04, Ke07}) or on temporal signals
(e.g. \cite{Ca06,Br06,Da09}). The link between these is novel.
On the other hand, in the
statistical physics of systems close to a phase transition, the
connection between spatial and temporal phenomena -like
hysteresis, critical slowing down, long range order, etc.-
is well known. While the theory of phase transitions is well understood in
thermodynamic equilibrium, its use in nonequilibrium systems is rather new.
However, many of the fundamental concepts of equilibrium phase transitions -
like scaling and universality- still apply in systems without a hermitian
Hamiltonian but rather defined by transition rates, for which the local
time-reversal symmetry is broken \cite{Od04}.Moreover, in nonequilibrium
systems a (dynamical) scaling of variables may occur even in first order
transitions, when the order
parameter jumps at the transition. This is exactly the kind of
phenomenon we are observing for a spatial ecological model.

\ack{We wish to thank V. Dakos, R. Donangelo, N. Mazzeo, M. Scheffer and E.
van Nes for many fruitful discussions. Work supported in part by PEDECIBA
(Uruguay) and Project PDT 63-013.}

\section*{References}
\bibliographystyle{unsrt}
\bibliography{ecosys}

\begin{thebibliography}{10}

\bibitem{Mc99}
L~J McCook.
\newblock Macroalgae, nutrients and phase shifts on coral reefs: scientific
  issues and management consequences for the great barrier reef.
\newblock {\em Coral Reef}, 18:357--367, 1999.

\bibitem{Ny00}
M~Nystrom et~al.
\newblock Coral reef disturbance and resilience in a human-dominated
  environment.
\newblock {\em Trends Ecol. Evol.}, 15:413--417, 2000.

\bibitem{Sc03}
M~Scheffer et~al.
\newblock Floating plant dominance as a stable state.
\newblock {\em Proc. Natl. Acad. Sci. USA}, 100:4040--4045, 2003.

\bibitem{Lu97}
D~Ludwig et~al.
\newblock Sustainability, stability, and resilience.
\newblock {\em Conservation Ecology}, 1(7), 1997.
\newblock http://www.consecol.org/vol1/iss1/art7.

\bibitem{Wa93}
B~H Walker.
\newblock Rangeland ecology: understanding and managing change.
\newblock {\em Ambio}, 22:2--3, 1993.

\bibitem{Sc98}
M~Scheffer.
\newblock {\em Ecology of Shallow Lakes}.
\newblock Chapman \& Hall, 1998.

\bibitem{Ca99}
S~R Carpenter et~al.
\newblock Management of eutrophication for lakes subject to potentially
  irreversible change.
\newblock {\em Ecol. Appl.}, 9:751--771, 1999.

\bibitem{Sc01}
M~Scheffer et~al.
\newblock Catastrophic shifts in ecosystems.
\newblock {\em Nature}, 413:591--596, 2001.

\bibitem{Carp01}
S~Carpenter.
\newblock Alternate states of ecosystems: evidence and some implications.
\newblock In Huntly~N. Press, M.~C. and S.~Levin, editors, {\em Ecology:
  achievement and challenge}, pages 357--381. Blackwell, London, UK, 2001.

\bibitem{St97}
E~K Steinberg and P~Kareiva.
\newblock Challenges and opportunities for empirical evaluation of spatial
  theory.
\newblock In D.~Tilman and P.~Kareiva, editors, {\em Ecology: achievement and
  challenge}, pages 318--332. Princeton University, 1997.

\bibitem{Le97}
S~A Levin and S~W Pacala.
\newblock Theories of simplification and scaling of spatially distributed
  processes.
\newblock In D.~Tilman and P.~Kareiva, editors, {\em Ecology: achievement and
  challenge}, pages 271--296. Princeton University, 1997.

\bibitem{Ag99}
M~R Aguiar and O~E Sala.
\newblock Patch structure, dynamics and implications for the functioning of
  arid ecosystems.
\newblock {\em Tree}, 14:273--277, 1999.

\bibitem{Kl99}
C~A Klausmeier.
\newblock Regular and irregular patterns in semiarid vegetation.
\newblock {\em Science}, 284:1826--1828, 1999.

\bibitem{Ha01}
J~von Hardenberg et~al.
\newblock Diversity of vegetation patterns and desertification.
\newblock {\em Phys. Rev. Lett.}, 87:1981011--1981014, 2001.

\bibitem{Ri04}
M~Rietkerk et~al.
\newblock Self-organized patchiness and catastrophic shifts in ecosystems.
\newblock {\em Science}, 305:1926--1929, 2004.

\bibitem{Ke07}
S~K\'efi et~al.
\newblock Spatial vegetation patterns and imminent desertification in
  mediterranean arid ecosystems.
\newblock {\em Nature}, 449:213--217, 2007.

\bibitem{Ca06}
S~R Carpenter and W~A Brock.
\newblock Rising variance: a leading indicator of ecological transition.
\newblock {\em Ecology Letters}, 9:311--318, 2006.

\bibitem{Br06}
W~A Brock and S~R Carpenter.
\newblock Variance as a leading indicator of regime shift in ecosystem
  services.
\newblock {\em Ecology and Society}, 11(9), 2006.
\newblock http://www.ecologyandsociety.org/vol11/iss2/art9/.

\bibitem{No75}
I~Noy-Meir.
\newblock Stability of grazing systems: an application of predator-prey graphs.
\newblock {\em Jour. of Ecology}, 63:459--482, 1975.

\bibitem{May77}
R~M May.
\newblock Thresholds and breakpoints in ecosystems with a multiplicity of
  stable states.
\newblock {\em Nature}, 269:471--477, 1977.

\bibitem{Th75}
R~Thom.
\newblock {\em Structural Stability and Morphogenesis}.
\newblock Reading: Benjamin, 1975.

\bibitem{VN05}
E~van Nes and M~Scheffer.
\newblock Implications of spatial heterogeneity for catastrophic regime shifts
  in ecosystems.
\newblock {\em Ecology}, 86:1797--1807, 2005.

\bibitem{Lu78}
D~Ludwig, D~D Jones, and C~S Holling.
\newblock Qualitative analysis of insect outbreak systems: the spruce budworm
  and forest.
\newblock {\em Jour. Anim. Ecology}, 47:315--332, 1978.

\bibitem{Mu93}
J~D Murray.
\newblock {\em Mathematical Biology}.
\newblock Springer-Verlag, 1993.

\bibitem{Hol59}
C~S Holling.
\newblock The components of predation as revealed by a study of small mammal
  predation of the european pine sawfly.
\newblock {\em Can. Entomol.}, 91:293--320, 1959.

\bibitem{Gil81}
R~Gilmore.
\newblock {\em Catastrophe Theory for Scientists and Engineers}.
\newblock Dover, 1981.

\bibitem{Pre07}
W~H Press et~al.
\newblock {\em Numerical Recipes. The Art of Scientific Computing}.
\newblock Cambridge University Press, third edition, 2007.

\bibitem{DOd06}
P~D'Odorico, F~Laio, and L~Ridolfi.
\newblock A probabilistic analysis of fire-induced tree-grass coexistence in
  savannas.
\newblock {\em The American Naturalist}, 167:E79--E87.

\bibitem{Pas02}
M~Pascual, M~Roy, F~Guichard, and G~Flierl.
\newblock Cluster size distributions: signatures of self-organization in
  spatial ecologies.
\newblock {\em Phil. Trans. R. Soc. Lond. B}, 357:657--666, 2002.

\bibitem{Van08}
J~Vandermeer, I~Perfecto, and S~M Philpott.
\newblock Clusters of ant colonies and robust criticality in a tropical
  agroecosystem.
\newblock {\em Nature}, 451:457--460, 2008.

\bibitem{Do09}
R~Donangelo, H~Fort, V~Dakos, M~Scheffer, and E~H van Nes.
\newblock Early warnings of catastrophic shifts in ecosystems: Comparison
  between spatial and temporal indicators.
\newblock {\em Int. Jour. Bif. and Chaos}, 2009.
\newblock at press.

\bibitem{Da09}
V~Dakos, M~Scheffer, E~H van Nes, V~Brovkin, V~Petoukhov, and H~Held.
\newblock Slowing down as an early warning signal for abrupt climate change.
\newblock {\em Proc. Natl. Acad. Scien.}, 105:663--724, 2008.

\bibitem{Od04}
G~\'Odor.
\newblock Universality classes in nonequilibrium lattice systems.
\newblock {\em Rev. Mod. Phys.}, 76:663--724, 2004.

\end{thebibliography}

\end{document}